\documentclass[twocolumn,superscriptaddress,showkeys,prl]{revtex4-2}
\usepackage{amsmath}
\usepackage{amssymb}
\usepackage{graphicx}
\usepackage{float}
\usepackage{url}
\usepackage{bm}
\usepackage{hyperref}
\usepackage[mathlines]{lineno}

\begin{document}

\title{Hyperuniform organization in human settlements}
\author{Lei Dong}
\email{leidong@pku.edu.cn}

\affiliation{Institute of Remote Sensing and Geographical Information Systems, School of Earth and Space Sciences, Peking University, Beijing 100871, China}
\date{\today}

\begin{abstract}
Quantifying the spatial organization of human settlements is fundamental to understanding the complexity of urban systems. However, the quantitative patterns of the distribution of villages, towns, and cities that lie between random and regular, are still largely unknown. Here, by analyzing the geographic location of settlements in diverse regions, we show that the apparently complex urban systems can be characterized by disordered hyperuniformity (with small density fluctuations), an intriguing pattern that has been identified in many physical and biological systems, but has rarely been documented in socio-economic systems. By introducing the mechanisms of spatial matching and competition, we develop a growth model that shows how settlements evolve towards hyperuniformity. Our model also predicts the heavy-tail population distribution across settlements, in agreement with empirical observations. These results provide insights into the self-organization of cities, and reveal the universality of spatial organization shared by social, physical, and biological systems.

\end{abstract}
\maketitle
Understanding the spatial organization of human settlements (e.g., villages, towns, and cities) is at the core of human geography and all urban studies. More generally, quantifying the spatial pattern of the key elements within a complex system is fundamental to understanding its dynamics. However, unlike physical systems or biological systems, whose dynamics have been extensively studied, there are very few quantitative models that can explain the spatial organization of settlements in urban systems. The most influential to date is still the Central Place Theory (CPT) proposed by the geographer Walter Christaller nearly a century ago~\cite{christaller1933zentralen}. According to CPT, settlements function as `central places' and tend to form in a two-dimensional (2D) hexagonal lattice because this configuration allows for optimal coverage of service areas. Under this hexagonal lattice hypothesis, the spatial structure of the urban system would be like crystals or some animal territories~\cite{barlow1974hexagonal}. 

\begin{figure}[htbp]
    \centering
    \includegraphics[width = 0.45\textwidth]{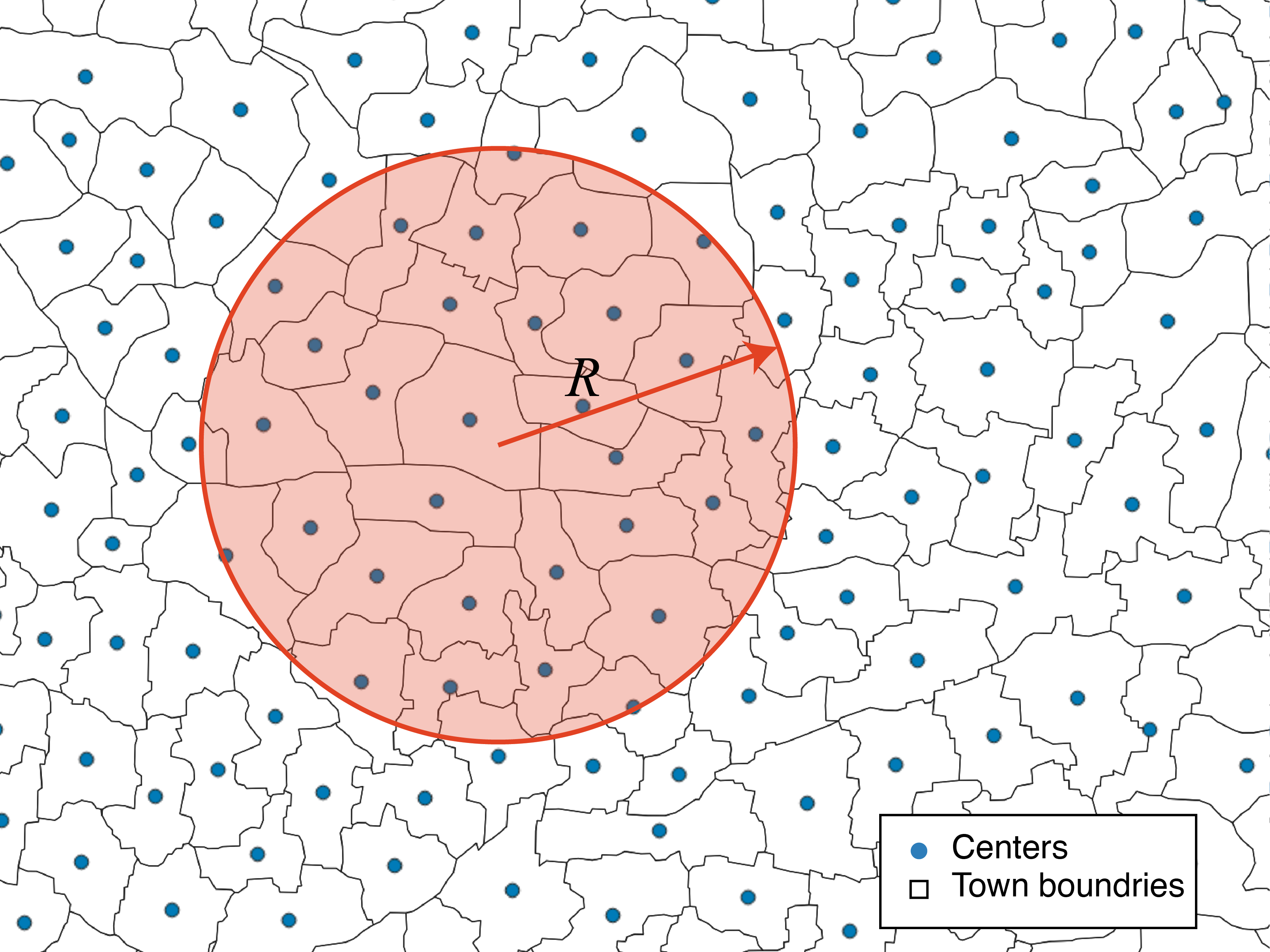}
    \caption{Spatial distributions of towns in China. The number of towns within the sampling window will fluctuate as the window radius and position vary. }
    \label{fig1}
\end{figure}

However, the spatial arrangement of settlements is often distorted by geographic constraints, and a perfect hexagonal structure does not exist in real urban systems. Although not as regular as a lattice, the arrangement of settlements is by no means random. In general, human settlements are located \textit{neither too close together nor too far apart} to effectively provide socioeconomic services to the surrounding areas (Fig. \ref{fig1}). Such a `disordered' pattern between randomness and regularity, as shown by many studies from biology~\cite{jiao2014avian} to physics~\cite{gabrielli2002glass,torquato2003local,weijs2015emergent,lei2019nonequilibrium}, mathematics~\cite{erdHos2013spectral}, and materials science~\cite{xie2013hyperuniformity,chremos2018hidden}, can be characterized by hyperuniformity, a point configuration with small density fluctuations at large distances. In biology, for example, the color-sensitive cone cells in chicken eyes exhibit hyperuniformity~\cite{kram2010avian}, which is thought to be the result of an optimization process during the evolution~\cite{jiao2014avian}. Many similar findings have been documented in natural systems (e.g., quasicrystals, galaxy clusters, active matter)~\cite{torquato2018hyperuniform,torquato2021swimming,PhysRevX.13.011038,klatt2019universal,oppenheimer2022hyperuniformity,huang2021circular}, but little is known about hyperuniformity in social systems in which humans interact and participate. In a previous work, researchers had found that the arrival time of buses in Cuernavaca City, Mexico, exhibits hyperuniform characteristics as a result of competition and mutual interaction among the drivers~\cite{krbalek2000statistical}, but that study analyzed only the time distribution and did not consider the 2D space.

Here, based on the analysis of geographic data in three highly diverse regions (China, Germany, and UK), we provide evidence showing that the spatial organization of settlements in an urban system -- a typical socio-economic system -- exhibits remarkable `effective hyperuniformity.' Note that since hyperuniformity is defined in an asymptotic limit (detailed later), it is hard to strictly obey the hyperuniform distribution for a real system, and the size (hundreds of cities) of the urban system is relatively small. Therefore, we use the term `effective hyperuniformity' for practical purposes~\cite{klatt2019universal}. To illustrate the origin of the observed pattern, we introduce two mechanisms, spatial matching and competition, to model the growth of settlements in an urban system. This model reveals the structural transitions of urban systems from randomness to regularity, and shows that the hyperuniformity of settlements arises during the formation process, rather than being rearranged after formation, as many existing models assume. It also predicts the heavy-tail population distribution across settlements, in agreement with empirical observations. Our results shed light on the long-standing central place conjecture in human geography and reveal the hidden universal patterns among social, physical, and biological systems.

\begin{figure*}[ht]
    \centering
    \includegraphics[width = 1.0\textwidth]{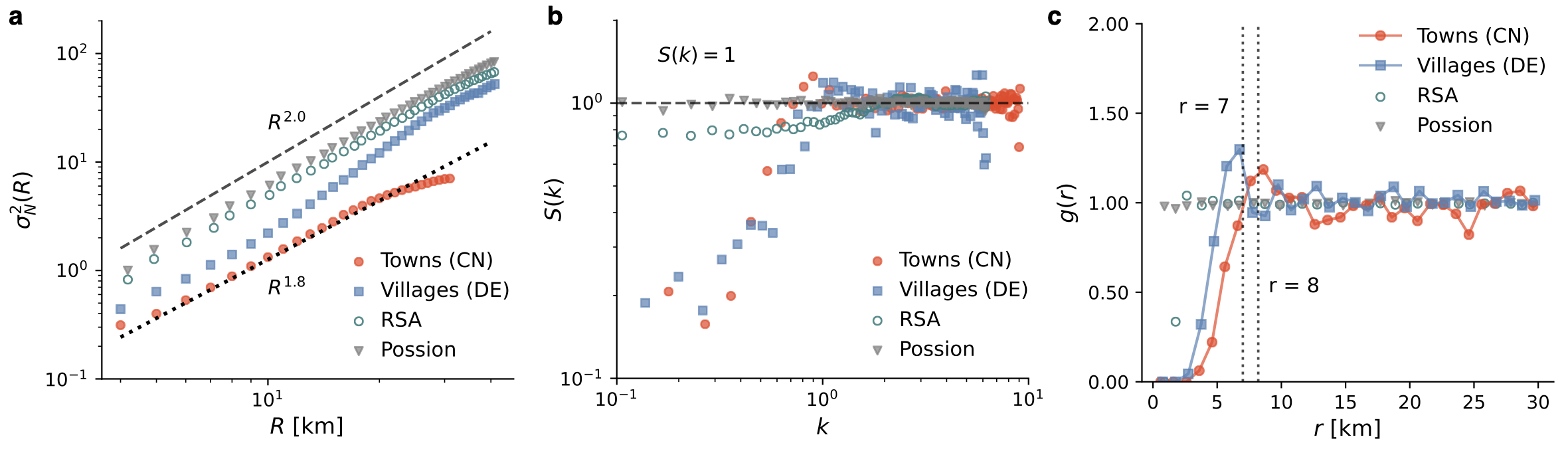}
    \caption{The number variance, structure factor, and pair-correlation function of the spatial structure of settlements in China (CN) and Germany (DE). (a) Number variance $\sigma_{N}^2$ as a function of the sampling window radius $R$. (b) Structure factor $S(k)$ as a function of the wave vector length $k$. (c) Pair correlation function $g(r)$ as a function of the distance $r$. The results of China were obtained by averaging 9 regions, and the results for each region are shown in SM. We also simulated the Poisson and RSA configurations 50 times (with the same $N$ and $L$ as the German data) and presented the averaged results. }
    \label{fig2}
\end{figure*}

\section*{Results}

\subsection{Effective hyperuniformity in diverse regions}
We first introduce the basic concept and measures of hyperuniformity. As a geometric concept, hyperuniformity characterizes the variance of the local number of points~\cite{torquato2018hyperuniform}. Let $R$ be the radius of a local sampling window (Fig.~\ref{fig1}), a point configuration is hyperuniform if the number variance of points $\sigma^2(R)$ within the window grows more slowly than the volume of the window (i.e., $\pi R^2$ in 2D), which can be characterized by three classes:

\begin{equation}
  \sigma_{N}^{2}(R) \sim 
    \begin{cases}
      R^{d-1} & \alpha > 1 \;\; \text{(CLASS I)}\\
      R^{d-1} \ln R & \alpha = 1 \;\; \text{(CLASS II)}\\
      R^{d-\alpha} & \alpha < 1 \;\; \text{(CLASS III)},
    \end{cases}
\label{number}
\end{equation}

\noindent where $d$ is the dimension of the system and $\alpha$ is a scaling exponent that related to the structure factor $S(k)$. Examples of Class I hyperuniform systems include all perfect crystals; Class II systems include some quasicrystals, perfect glasses, and the early universe; and Class III includes random organization models, classical disordered ground states, and perturbed lattices~\cite{torquato2018hyperuniform}, etc. For hyperuniform systems, the structure factors exhibit a scaling law $S(k) \sim k^\alpha$, and $S(k)$ diverges to $0$ as the wave number $k \rightarrow 0$, implying that the density fluctuations of the system are diminished for small values of $k$~\cite{torquato2018hyperuniform}. Given a point configuration, the structure factor $S(k)$ can be calculated by the equation $S(\textbf{k}) = \frac{1}{N}|\sum_{i}^{N} \text{exp} (i \textbf{k} \cdot \textbf{r}_i)|^2$, where $N$ is the number of points in the system, $\textbf{r}_i$ is the position of point $i$, and $\textbf{k}$ is the wave vector~\cite{hexner2017enhanced}.

To quantify the spatial structure and demonstrate the emergence of hyperuniformity in urban systems, we collected three datasets: the township boundaries in China in 2020, the villages (\textit{Hauptdorf}) in Germany in the 1930s, and the Middle Super Output Area (MSOA) in London (UK) in 2011. The village data for Germany were derived by manually digitizing Christaller's original map \cite{christaller1941zentralen}. These datasets characterize the distribution of settlements in three very different countries and span a period of almost one hundred years, which helps us to test the robustness of our results. For the Chinese and UK data, we compute the geometric centers of the township/MSOA administrative boundaries and use these centers to analyze the spatial pattern. For the German data, we directly used the point locations of the villages in the original map. To compute the Euclidean distance between points, we convert the latitude and longitude coordinates into a projected coordinate system, and select a square area of $L \times L$ for further analysis ($L = 200 km$ in Germany, $L=100 km$ in China, and $L = 20 km$ in London), see Methods and Supplementary Fig. 1-2 for details.

First, we compute the number variance $\sigma^2_{N}(R)$. As shown in Fig.~\ref{fig2}a and Supplementary Fig. 3, the settlement number exhibits low fluctuations, and the slopes of the fitting lines satisfy the Class III hyperuniformity $\sigma_{N}^{2}(R) \sim R^{d-\alpha}$. More interestingly, we find that the number fluctuation of settlements in Germany exhibits effective hyperuniformity at short length scales (small $R$s), while it is closer to the Poisson pattern at large scales (for large $R$s, the number fluctuation increases and the scaling exponent $\sim 2$). This finding that short- and long-range scales in a system exhibit different number and density fluctuations is also consistent with some theoretical models~\cite{hexner2015hyperuniformity,hexner2017enhanced} and empirical studies~\cite{zhang2022hyperuniform}. For example, a recent theoretical work has shown that the point configuration generated by the random organization model, for the absorbing state just reaching the threshold, the observed hyperuniformity at short scales becomes closer to the Poisson pattern at large scales~\cite{hexner2017enhanced}.

The transition from a hyperuniform regime to large variance fluctuations takes place at length scales corresponding to the characteristic size of the urban system, which is reflected by the peak in the pair correlation function $g(\textbf{r})$. $g(\textbf{r})$ measures how the point density $\rho$ varies as a function of the radial distance $r \equiv |\textbf{r}|$ from a reference point. In a two-dimensional space, $g(r) \sim \frac{n(r, r+dr)}{\rho 2\pi r dr}$, where $n(r, r+dr)$ is the number of points within a distance of $r$ and $r+dr$ away from a point and $\rho$ is the number of points per unit area. Visually, Fig. 1 shows that each settlement occupies an effective exclusive region, making the cities be separated from each other as much as possible. The characteristic size of this exclusive region is captured quantitatively by the pronounced peak in $g(r)$. As we can see from Fig.~\ref{fig2}c, the peak of $g(r)$ corresponds exactly to the average nearest neighbor distance of settlements ($\langle r^{\text{DE}} \rangle = L/\sqrt{N} \approx 7 km$ and $\langle r^{\text{CN}} \rangle \approx 8 km$). Note that this distance corresponds to a travel time of $\approx 2$ hours on foot ($10$-$15$ min by car), making it easy for settlements to reach each other, but keeping enough distance to avoid competition.

\begin{figure*}[htp]
    \centering
    \includegraphics[width = .9\textwidth]{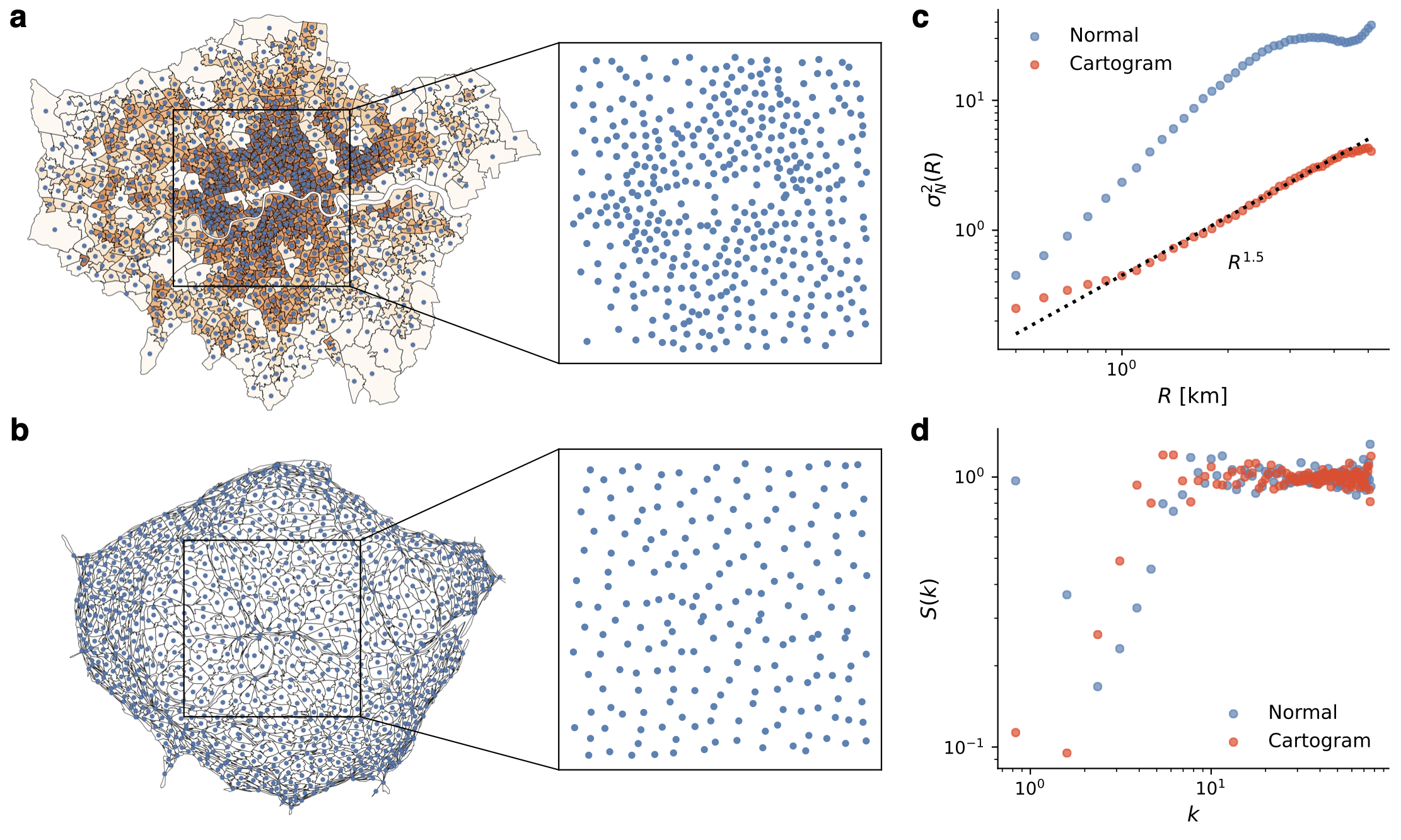}
    \caption{Spatial patterns of urban settlements in London. (a) Blue dots represent the center of the Middle Super Output Area (MSOA). Orange-colored population densities are from the 2011 UK Census. (b) The density-equalizing projection. The associated number variances (c) and structure factors (d).}
    \label{fig3}
\end{figure*}

The evidence of effective hyperuniform structure in urban systems can also be verified by analyzing of the structure factor $S(k)$. For computational purposes, we use the expanded structure factor equation according to Ref.~\cite{zhang2016concept}, $S(\textbf{k}) = \frac{1}{N} \langle |\sum_{i}^{N} \text{cos} (\textbf{k} \cdot \textbf{r}_i)|^2 + |\sum_{i}^{N} \text{sin} (\textbf{k} \cdot \textbf{r}_i)|^2\rangle$. We generate wave vectors at values of $\textbf{k} = (i, j) \frac{2\pi}{L}$, where $i$ and $j$ are integers from $1$ to $L$. Similar to the results of the number variance and the pair correlation function, the structure factors $S(k) \rightarrow 0 $ for $k \rightarrow 0$ (Fig.~\ref{fig3}b). Because the number of settlements in the urban system is relatively small, the data we can observe are not sufficient to fit a robust scaling law exponent under the small $k$ setting. However, using the number variance and the Class III hyperuniformity properties, we can approximate that $\alpha \approx 0.1$ for Germany and $0.2$ for China (Fig.~\ref{fig2}a). For comparison, we also plot the results of Poisson and random-sequential-adsorption (RSA) configurations~\cite{evans1993random} for the same number of points and system size. Figure~\ref{fig2}b shows that Possion and RSA systems do not have the small $k$ effect, and the pair correlation function of the settlements becomes 0 when $r \rightarrow 0$, which is different from the Poisson systems (Fig. \ref{fig2}c).

Notably, the spatial dispersion pattern of urban settlements is influenced by natural and socio-economic factors, and therefore in most cases does not appear to be hyperuniform. In particular, urban systems have the agglomeration effect, which leads to closer urban clusters at the regional level and the concentration of citizens within cities~\cite{duranton2004micro}. In other words, the urban system is not a homogeneous system that previous hyperuniform studies focus on. To address this challenge, we draw inspiration from some existing work on space deformation (e.g., the density-equalizing projection, also known as the cartogram) to reproject the map using population density~\cite{tobler2004thirty,gastner2004diffusion,gastner2006optimal}. In the physical sense, it can be thought of calculating an \textit{effective distance} using the population density as weights. In the socioeconomic sense, the area covered by the facility after the projection directly corresponds to the size of the population served. Since the agglomeration effect is more significant within cities than at the regional level, we use settlements within London (UK) as an example to show the spatial patterns before and after the density-equalizing projection (Fig.~\ref{fig3}ab, see Methods for details). Figure~\ref{fig3} demonstrates that even with very high heterogeneity, the spatial distribution of settlements after reprojection with population density conforms well to the hyperuniformity with $\sigma_{N}^{2}(R) \sim R^{1.5}$ and $S(k) \rightarrow 0$ when $k$ is small.

\subsection{The matching growth model}
\begin{figure*}[ht]
    \centering
    \includegraphics[width = 1.\textwidth]{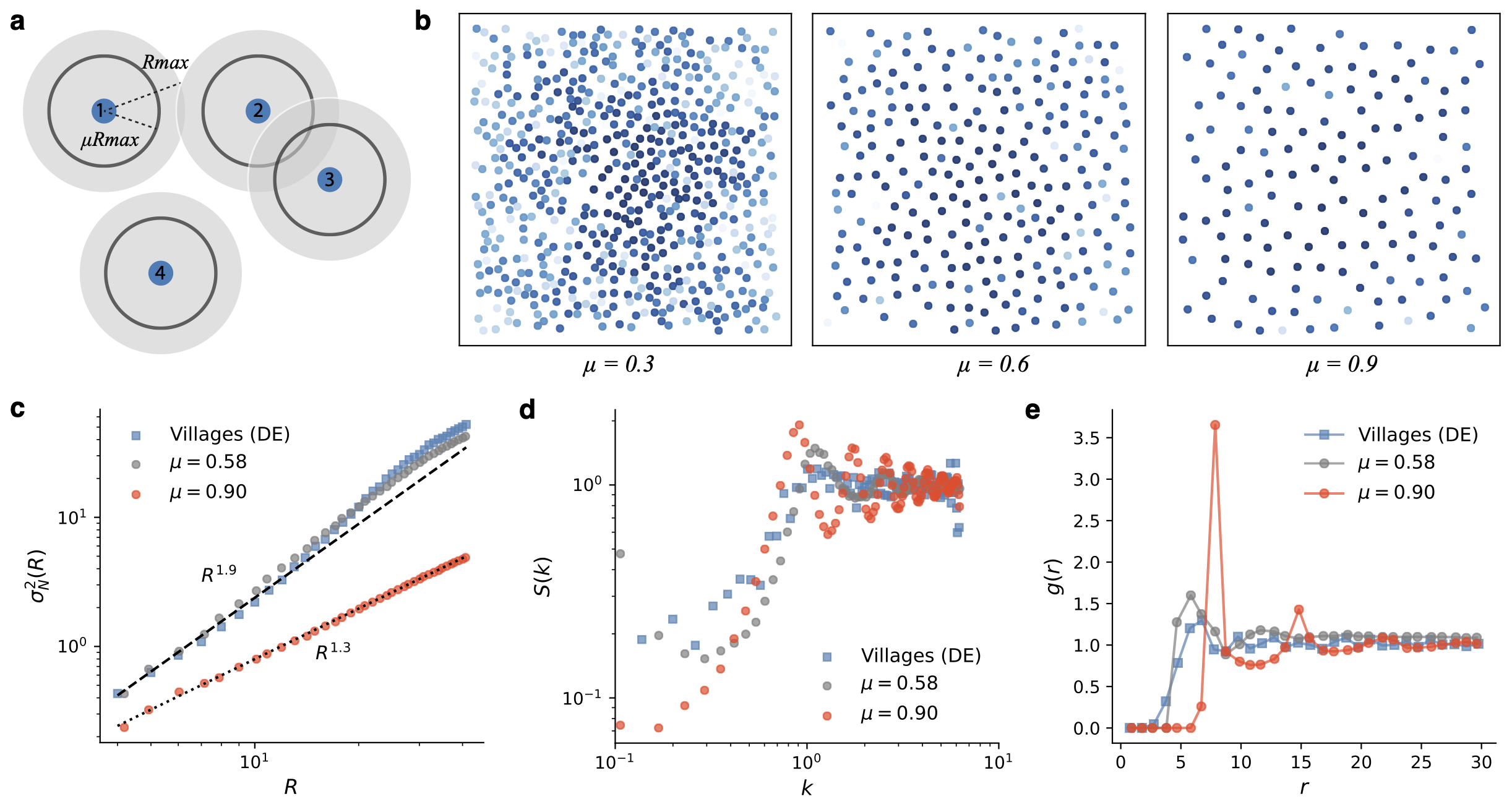}
    \caption{Illustration of the model and simulation results. (a) The index numbers of the settlements are the order in which they were added. The large gray disks with radius $R_{max}$ around the settlements are their interaction regions. A settlement can only survive if its interaction area (gray disk) overlaps with that of the existing settlements. Therefore, settlement 4 cannot survive. Meanwhile, a settlement will merge with existing settlements if its distance to the nearest point is less than $2\mu R_{max}$. Thus, settlement 3 will merge with settlement 2. (b) The point configurations generated by the model. The color ranges from dark to light representing the time of appearance from early to late. (c-e) Comparison of the statistics of the empirical results (German villages) and the simulated results. (c) Number variances. (d) Structure factors. (e) Pair correlation functions. In the simulation, we set $R_{max} = 4$, and for each value of $\mu$, we generated 50 point configurations and averaged the statistics.}
    \label{fig4}
\end{figure*}

What causes urban systems to become hyperuniform? Although several models (e.g., the Manna model, the random organization model~\cite{hexner2017enhanced}, and the Lloyd algorithm~\cite{klatt2019universal}) have been developed to explain the emergence of hyperuniformity in point configurations, these models tend to fix the number of points in the system, which is inconsistent with the fact that the development of settlements in an urban system is a growing process. Moreover, existing models generate hyperuniformity by allowing points to move randomly, whereas in an urban system, it is difficult for a settlement to change its spatial position once it is formed. Thus, we conjecture that the hyperuniformity of urban systems arises during the process of settlement formation, rather than being rearranged after formation. To simulate such a process, we propose a matching growth model with two simple assumptions: 1) spatial matching attachment; 2) a minimum distance to avoid competition. The the model is illustrated in Fig.~\ref{fig4}a.

Specifically, assume that human settlements grow in a two-dimensional Euclidean space where resources are uniformly distributed. At $t=0$ a first settlement appears in the center of the space. This settlement begins to attract people and provide services to the surrounding area. After a certain time step $t$ a new settlement  $i$ appears (each newly generated settlement has a constant population size). Since each settlement has a limited interaction radius $R_{max}$, the newly added settlement cannot be too far away from existing settlements or it will not survive (this matching growth mechanism was also used in Refs.~\cite{zhang2015scaling,li2017simple}). Similarly, due to limited space and resources, the new settlement $i$ cannot be too close to existing settlements, or it will merged with the nearby settlements $j$s with probability $f_j$. The probability of which settlement it will merge with is proportional to the population size of that settlement. This merging mechanism can be seen as a local preferential attachment in complex networks~\cite{barabasi2009scale}, and it will only affect the population size of settlements and will not affect their spatial distribution or the emergence of hyperuniformity. Therefore, the conditions for a settlement $i$ to survive are satisfied by

\begin{equation}
2\mu R_{max} \leq |\textbf{r}_i - \textbf{r}_j| \leq 2R_{max},
\end{equation}

\noindent where $\textbf{r}_i$ is the location of the settlement $i$, $\textbf{r}_j$ ($j$ from $1$ to $N$, $j \neq i$) are the locations of existing settlements, $|\cdot|$ is the Euclidean distance, and $\mu$ ($0 \leq \mu \leq 1$) is the key parameter in our model. If $\mu \rightarrow 1$, the settlements would be arranged close to a lattice; if $\mu \rightarrow 0$, the process becomes a random matching growth, producing a point density distribution that decays from the center to the periphery~\cite{zhang2015scaling,li2017simple}.

Figure~\ref{fig4} shows that this simple model accurately reproduces the hyperuniform characteristics of the urban system, as well as the dynamic and structural transitions from randomness to regularity. The generating configuration shown in the middle panel of Fig.~\ref{fig4}b is visually almost indistinguishable from the actual urban settlement patterns. We further quantify the statistical metrics and present the results in Fig.~\ref{fig4}c-e. The simulation results show that the associated number variance, the structure factor, and the pair correlation functions are in good agreement with the empirical data when $\mu = 0.58$. The good fit of the model simulations strongly suggests that the effective hyperuniform patterns in urban systems do indeed arise from the matching and competition between settlements, implying that there is a strong bottom-up optimization process in the formation of settlements. Importantly, our model also captures the transition structure at long-range scales, i.e. for large $Rs$ the number variances become closer to the Poisson pattern (Fig.~\ref{fig4}c).

\begin{figure*}[ht]
    \centering
    \includegraphics[width = .75\textwidth]{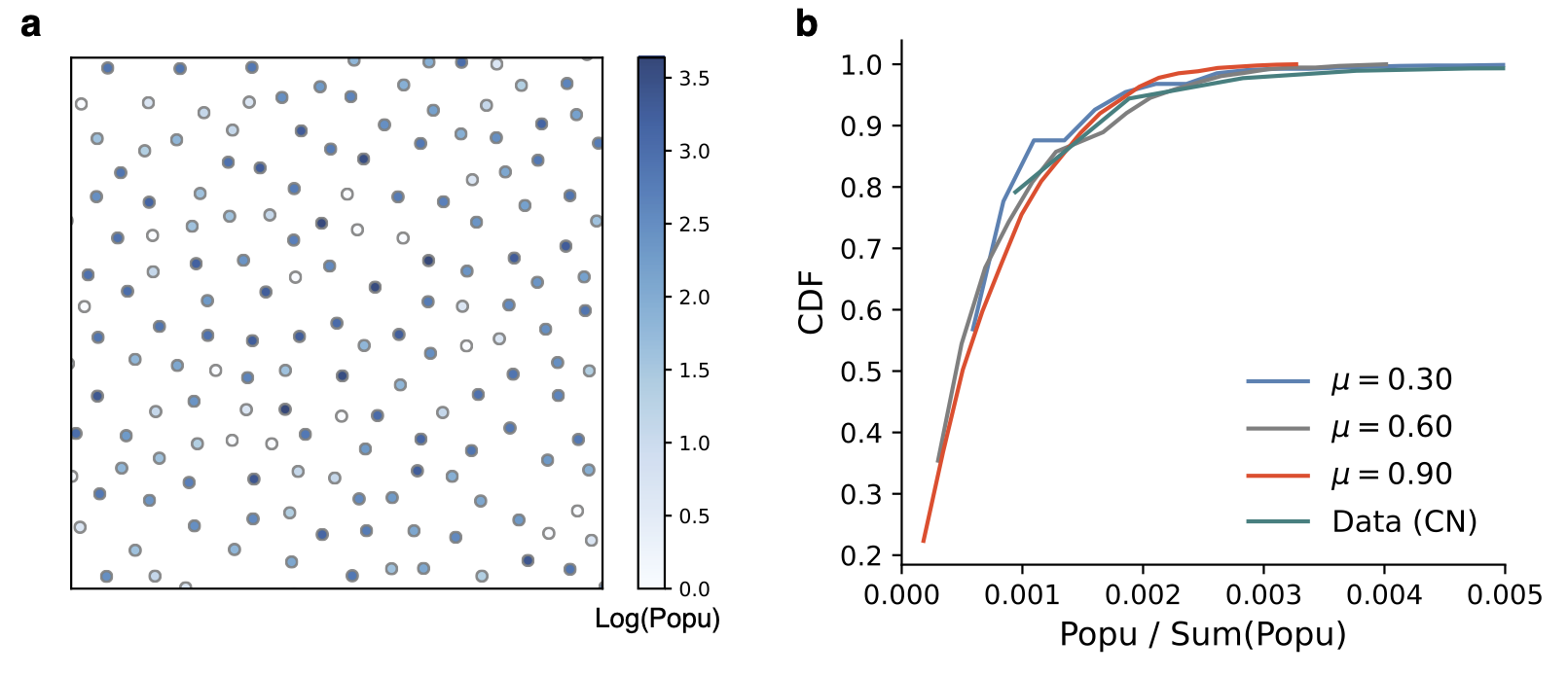}
    \caption{Model predicted population distribution. (a) The spatial structure of settlements colored by population size ($\mu = 0.6$, $R_{max} = 5$). (b) The cumulative density function (CDF) of the population. To make the data comparable between simulations and empirical data, we divided the population of each settlement by the total population in the system (thus ensuring that the total population is the same for each curve). In the simulation, we set $R_{max} = 5$, and for each value of $\mu$, we generated 50 point configurations and averaged the statistics. The simulation stops when the number of settlements reaches 1,500.}
    \label{fig5}
\end{figure*}

Another key and unique empirical observation in urban systems is population heterogeneity: there are many small towns, but few big cities. By accounting for population growth, our model can predict such heterogeneous population distribution across settlements. Figure~\ref{fig5}a illustrates the population size of each settlement by color, with darker colors representing larger populations in a logarithmic scale. Quantitatively, in Fig.~\ref{fig5}b, we show the cumulative distribution function (CDF) of the population with different $\mu$s, and it can be seen that the CDF fits well with the population data of Chinese towns (calculated using Worldpop data~\cite{tatem2017worldpop}). The CDF curves illustrate that the larger the settlements are in population size, the fewer number they will be. These results further suggest that our growth model can not only generate a hyperuniform point configuration, but also reproduce the point weights (i.e., population).

\section*{Discussion}

Given the many complexities that influence human settlements -- heterogeneous distribution of services, geography, culture, etc. -- one would not expect something as simple as hyperuniformity to emerge. This revealed hidden pattern advances our understanding of urban systems, and provides quantitative support for urban theories, such as the well-known CPT. Compared to the strong lattice assumptions of CPT, the hyperuniform tessellation is closer to the real urban pattern. To understand the origin of hyperuniformity in urban systems, we propose a simple growth model that suggests that the spatial matching and competition lead to its emergence. Our model overcomes the shortcomings of the CPT (and some other urban models), which do not consider the temporal aspect of settlement formation, and may have the potential to be used in biological systems or ecosystems where spatial growth and competition are important mechanisms governing system dynamics. 

We also note that the model is not designed at a specific spatial scale, and we mainly compare the simulation results with empirical data at the regional level. When zooming into the inner city, the model can also simulate the decay curve of population density from the center to the periphery, an important measure of urban agglomeration and population distribution. Supplementary Fig. 4 shows that the population distribution within the city is more concentrated when $\mu$ is small (more random), and more dispersed when $\mu$ is large (more regular), and the density decay curve is close to the `crater pattern' mentioned in previous studies~\cite{newling1969spatial,mcdonald1989econometric} -- the population density first increases and then decreases as the distance from the city center increases.

Finally, from an application point of view, the hyperuniformity might be used as a tool for planning and/or optimizing the layout of facilities within cities. For example, a question worth investigating for urban planning is: How efficient is the distribution of our settlements? Or is the spatial organization of settlements optimized? Similar to the particle packing, given the minimum distance and the number of settlements, we can calculate the packing fraction $\phi$ of human settlements: $\phi = \frac{1}{L^2} \sum_{i}^{N} \pi (\mu R_{max})^2$, where $N$ is the number of settlements in the $L \times L$ space. The packing fraction measures the fraction of space covered by particles. In urban systems, $\phi$ can be used to measure the efficiency of settlement tessellations. At $\mu = 0.58$, the resulting settlement configuration has a packing fraction of $\phi = 0.38$, and $\mu = 0.90$ corresponds to a $\phi = 0.56$, which is very close to the number of some biological systems~\cite{jiao2014avian}, suggesting that humans are able to achieve near-optimal results in group behavior.

\section*{Methods}

\subsection{China township data}
Settlement data for Chinese townships were obtained from online Internet maps. We calculated the population in each town by overlaying WorldPop's population distribution data~\cite{tatem2017worldpop} with the town's boundaries. To reduce the influence of geographical factors (e.g., rivers, mountains), we used nine 100km $\times$ 100km areas in the plains as the study area, which contains an average of 147 towns per snapshot (Supplementary Fig. 1). The number variance of each snapshot is shown in Supplementary Fig. 3. Note that townships are the fourth level of division in China, after provinces, cities, and counties. 

\subsection{Germany village data}

The settlement data for Germany were manually digitized from one of Christaller's maps \cite{christaller1941zentralen}. There are seven settlement classes (e.g., villages, counties, cities, and districts) in the original map, we used the two lowest classes: villages (\textit{Hauptdorf}) and high-level villages (\textit{Gehobenes Hauptdorf}). This is because these are more data points included in these two classes, which allows us to perform statistical analysis. There are a total of 794 villages within the 200km $\times$ 200km area we studied (Supplementary Fig. 2).

\subsection{London MSOA census data}

We collected the London census data from Ref.~\cite{UK} and use the centers of Middle Super Output Area (MSOA) boundaries as a proxy for settlements. By definition, MSOAs should cover roughly equal amounts of population (a few thousand), so they vary greatly in size: in denser areas MSOAs are spatially smaller and in suburban areas they are large. The advantages of using MSOAs here are twofold: on the one hand, the UK Census provides population data for MSOAs, allowing us to make the equal-density projection; on the other hand, the population size within an MSOA is relatively close to the size of a settlement. Fig.~3b in the main text shows the equal population density cartogram. We then selected a $20km \times 20km$ area (Inner London) of the city center for our analysis, which contains 430 and 196 settlements before and after the equal-density projection, respectively.

\subsection{Density-equalizing map projections and cartogram}

Cartograms (or density-equalizing map projects) are maps in which the geographic regions are rescaled in order to make areas are proportional to given variables (e.g., population size, economic output). During rescaling, cartograms can preserve the topological relationship of spatial units, resulting in both insightful visualizations and convenient statistical analysis~\cite{tobler2004thirty,gastner2004diffusion,gastner2006optimal,gastner2018fast}. In the urban setting, cartograms can help us transform a heterogeneous system into a homogeneous one, which is the focus of previous hyperuniform studies. And cartograms in urban systems have both physical and socioeconomic meanings. In the physical sense, the equal-density projection of population can be seen as a diffusion process, where people `diffuse' from high density areas to low density areas until the densities are equal, and the linear diffusion is also one of the main methods to derive the cartogram~\cite{gastner2004diffusion}. In the socio-economic sense, the area covered by the facility after the projection directly corresponds to the size of the population served. Here, we use the QGIS plugin implements the algorithm proposed by Ref.~\cite{dougenik1985algorithm} to create the cartogram.

\subsection{Acknowledgements}
We thank Y. Liu, Y. Wang, and J. Zhang for their help. L.D. acknowledge the support of the Fundamental Research Funds for the Central Universities, Peking University.

\subsection{Data and code availability}
All data necessary to reproduce our results are available through \url{https://github.com/leiii/hyperuniform-cities}.

\end{document}